\documentclass[a4paper,12pt]{article}
\usepackage{latexsym}
\usepackage{multirow}
\usepackage{graphics}
\usepackage{graphicx}
\usepackage[english]{babel}
\usepackage{natbib}
\usepackage{amssymb}
\usepackage{amsthm}
\usepackage{amsmath}
\usepackage{eurosym}
\usepackage{color,colortbl}
\usepackage{booktabs,caption}
\usepackage[flushleft]{threeparttable}
\usepackage[dvipsnames]{xcolor}
\usepackage[margin=3cm]{geometry}
\usepackage{url}
\usepackage{hyperref}

\renewcommand{\baselinestretch}{1.1}
\bibpunct[: ]{(}{)}{;}{a}{,}{;}
\title{\Huge{Do Losses Matter? The Effect of Information-Search Technologies on Risky Choices}\footnote{This version: {\itshape September 2023}. We would like to thank Marco Tecilla for superb computer programming. We are grateful to Miguel A. Mel\'{e}ndez-Jim\'{e}nez for helpful suggestions. We also thank the seminar audience at the University of Stirling for their comments. Any error is our own responsibility. Conflicts of interest: none.}}
\author{
Luigi Mittone\footnote{University of Trento (Italy). E-mail: luigi.mittone@unitn.it }\\
\and 
Mauro Papi\footnote{University of Aberdeen (UK). E-mail: m.papi@abdn.ac.uk (corresponding author)}\\
}
\date{}

\begin{document}

\renewcommand{\baselinestretch}{1.0}

\maketitle

\vspace{1cm}

\begin{abstract}
Despite its importance, relatively little attention has been devoted to studying the effects of exposing individuals to digital choice interfaces. In two pre-registered lottery-choice experiments, we administer three information-search technologies that are based on well-known heuristics: in the ABS (alternative-based search) treatment, subjects explore outcomes and corresponding probabilities {\itshape within} lotteries; in the CBS (characteristic-based search) treatment, subjects explore outcomes and corresponding probabilities {\itshape across} lotteries; in the Baseline treatment, subjects view outcomes and corresponding probabilities {\itshape all at once}. We find that (i) when lottery outcomes comprise gains and losses (experiment 1), exposing subjects to the CBS technology systematically makes them choose {\itshape safer} lotteries, compared to the subjects that are exposed to the other technologies, and (ii) when lottery outcomes comprise gains only (experiment 2), the above results are reversed: exposing subjects to the CBS technology systematically makes them choose {\itshape riskier} lotteries. By combining the information-search and choice analysis, we offer an interpretation of our results that is based on prospect theory, whereby the information-search technology subjects are exposed to contributes to determine the level of attention that the lottery attributes receive, which in turn has an effect on the reference point.
\end{abstract}


\vspace{0.2cm}

\newpage

\section{Introduction}

Choice architecture and nudging have successfully been utilised to improve people's decisions in various domains, such as health \citep{JohnsonGoldstein2003} and finance \citep{ThalerBenartzi2004}.\footnote{See \citet{ThalerSunstein2008} for a review.} At the same time, growing evidence has documented a widespread use of `sludges' or `dark patterns', especially in the online choice environment \citep{MathurAcarFriedmanLucheriniMayerChettyNarayanan2019,KozyrevaLewandowskyHertwig2020}.\footnote{\citet{MathurAcarFriedmanLucheriniMayerChettyNarayanan2019} define dark patterns as `{\itshape user interface design choices that benefit an online service by coercing, steering, or deceiving users into making unintended and potentially harmful decisions}'.} These refer to the phenomena whereby a choice architect, such as an online firm, designs the choice environment with the objective of exploiting the consumers' cognitive biases at its advantage, by manipulating the way information is presented, steering, using decoys, suitably  selecting the default option, etc.\footnote{We here use the term `choice architect' in broader sense than \citet{ThalerSunstein2008}, by considering a choice architect as an agent who designs a choice environment to generate an economic advantage (e.g., profit) for themselves, independently of the effects (positive or negative) that this can have on the decision maker.} 
Policy-makers, such as the UK Competition and Markets Authority (CMA), have acknowledged the benefits of online choice architecture, such as that of simplifying the overwhelming amount of information available, but - at the same time - expressed concerns about its potentially harmful effects and, in multiple occasions, taken actions \citep{CMA2020}.\footnote{For example, see \citet{CMA2017}.} 

Despite its importance, relatively little attention has been devoted to the study of the effects of exposing consumers to different online choice interfaces. As pointed out by \citet{WeinmannSchneiderVomBrocke2016}, research on `digital nudging' is still in its infancy phase and {\itshape `information systems designers must understand the behavioral effects of interface design elements so that digital nudging does not happen at random and unintended effects do not occur'}. A particularly important domain of applicability of digital nudging is that of financial decisions \citep{Benartzi2017}. In fact, it is well-known that (i) people typically struggle to understand probabilistic objects \citep{Benjamin2019}, and (ii) certain financial decisions, such as those pertaining to retirement savings and mortgages, are made infrequently, which limits the possibility of learning from mistakes \citep{ErtaHuntIscenkoBrambley2013}. Furthermore, the rapid expansion of digital technologies has enhanced the possibilities of creating even more sophisticated online choice interfaces \citep{Cai2020} that - if suitably designed - can possibly further exploit the behavioural biases observed in the offline environment. Thus, the purpose of this work is to investigate, in a controlled laboratory setting, whether administering different information-search technologies has an effect on people's decisions in a risky-choice setting. 

In two pre-registered experiments, we administer three information-search technologies that are based on well-known heuristics by suitably modifying the Mouse-Tracing paradigm \citep{PayneBettmanJohnson1993}: in the ABS (alternative-based search) treatment, subjects explore the attributes {\itshape within} alternatives; in the CBS (characteristic-based search) treatment, subjects explore the attributes {\itshape across} alternatives; in the Baseline treatment, subjects view the attributes {\itshape all at once}.\footnote{Theoretical examples of ABS and CBS procedures are expected-utility maximisation and lexicographic procedures, respectively \citep{PayneBettmanJohnson1993}.} Our design involves binary lotteries, which - although considerably simpler - can be thought of as an abstraction of financial products, such as loans, mortgages, and insurance contracts.\footnote{\citet{NiedduPandolfi2020} experimentally investigate the impact of financial literacy on people's willingness to invest by modelling financial assets as binary lotteries. They argue that the financial skills needed in their setup are key to make knowledgeable decisions in the real world.} In particular, every choice problem in these experiments comprises one risky and one safe lottery, with the property that every risky lottery is a mean-preserving spread of every safe lottery \citep{RothschildStiglitz1970}. The experiments are between-subject and treatments are identical, except for the technology subjects are administered to explore the lottery attributes.\footnote{In these experiments, the lottery attributes (or lottery characteristics) are given by the `high prize', the `probability that the high prize occurs', and the `low prize'.}  

Our contribution is three-fold. First, we identify an interplay between the technology subjects are exposed to and the lottery-outcome domain. In particular, we provide evidence that, when lottery outcomes comprise gains and losses (experiment 1), the subjects that are administered the CBS technology make relatively {\itshape safer} choices, compared to the subjects that are exposed to the ABS or are not induced to use any particular procedure (Baseline treatment). In contrast, when lottery outcomes comprise only gains (experiment 2), we obtain the opposite result: the subjects that are administered the CBS technology make relatively {\itshape riskier} choices. The lotteries used in experiment 2 are identical to those used in experiment 1, up to a summation to all lottery outcomes of the same positive constant. We also find that the ABS subjects and the subjects that are not to induced to use any procedure do not exhibit a different choice behaviour across lottery-outcome domains. 

The key aspect of our results is that we obtain preference reversals by simply exposing subjects to different information-search technologies, while {\itshape holding both the outcomes and the corresponding probabilities of the lotteries available constant}. Our second contribution is thus to propose an alternative rationalisation of our results that is based on prospect theory (PT). At the core of the matter is the combined effect on the formation of the reference point of, not only the lottery-outcome domain, but also the procedure subjects are induced to use. In particular, we argue that the information-search technology subjects are exposed to contributes to determine the level of attention that the lottery attributes receive, which in turn has an effect on the PT preference parameters.\footnote{Results were largely unexpected, and turned out to be more interesting than initially conjectured. As such, the proposed explanation has been derived {\itshape ex post}.}

In PT the reference point is typically given by a function of the lottery attributes \citep{BaillonBleichrodtSpinu2020}. Thus, by transitioning from the gain-loss lottery-outcome domain (experiment 1) to the gain lottery-outcome domain (experiment 2), the reference point is likely to increase. Since the subjects who are induced to use a CBS procedure view one outcome at a time, then the `high prize' becomes relatively more salient and - as a result - gains more weight in the formation of the reference point. This leads to a further rightward shift of the reference point in the gain domain. As such, the CBS subjects are more likely to perceive a given lottery outcome as a loss compared to the subjects assigned to the other treatments. Given that in the loss domain PT decision-makers are risk lovers \citep{KahnemanTversky1979}, CBS subjects tend to make riskier choices in the gain domain. 

On the other hand, the ‘holistic view’ that is enforced in the ABS and Baseline treatments - whereby subjects simultaneously view both outcomes and corresponding probabilities of each lottery - acts as a moderator in the formation of the reference point, which results in the other lottery attributes (other than the ‘high prize’) gaining relatively more importance. This implies that, when the lottery-outcome domain comprises gains and losses (experiment 1), the reference point tends to be relatively higher, and when the lottery-outcome domain comprises only gains (experiment 2), the reference point tends to be relatively smaller, compared to that of the CBS treatment. Thus, the subjects assigned to the ABS and Baseline treatments choose in a relatively riskier (resp., safer) way in experiment 1 (resp., experiment 2). 

In an earlier study \citep{MittonePapi2020} we find that, when lottery outcomes comprise only gains, inducing subjects to use a CBS heuristic makes them choose riskier options. The results of experiment 2 have been obtained by considering different lotteries than those used in \citet{MittonePapi2020} and recruiting from a different subject population. As such, our third contribution is that experiment 2 constitutes a replication of our earlier results by providing an additional robustness check of the identified phenomenon.

We conclude this section by discussing two wider implications of our results. First, in standard and behavioural models of risky choice, such as prospect theory \citep{KahnemanTversky1979} and salience theory \citep{BordaloGennaioliShleifer2012}, the way the decision-maker allocates attention is determined by the lottery attributes.\footnote{For example, in prospect theory, rare events are over-weighted \citep{KahnemanTversky1979}. On the other hand, in salience theory, a state of the world $\tilde{s}$ is more salient than some other state $s$ whenever $\tilde{s}$ stands out, i.e., the payoff range of state $s$ is contained in the payoff range of state $\tilde{s}$ \citep{BordaloGennaioliShleifer2012}.} In contrast, our evidence is in line with the principle whereby the attention that the decision-maker pays to the lottery attributes - which in these experiments is induced by the exposure to a certain procedure - is a primitive element of the decision process, and, as such, is an input (and not an output) of the decision process \citep{OrquinMuellerLoose2013}. By letting the reference point depend on, not only the lottery-outcome domain, but also the procedure subjects are induced to use, the proposed explanation of the results is consistent with this idea \citep{PachurSchultemecklenbeckMurphyHertwig2018,HirmasEngelmann2023}.

Second, this paper demonstrates that, in a controlled laboratory setting, the combined effect of the lottery-outcome domain and the administered information-search technologies can affect people's risky choices in specific ways. Therefore, our results are of significant importance for the design and the framing of real-world digital choice platforms \citep{ErtaHuntIscenkoBrambley2013,Cai2020,CMA2020}. For example, suppose that an online financial firm prefers customers to purchase `risky mortgages' because they are more profitable. According to our results, one way of achieving the desired objective would be for the firm under consideration to frame the mortgage characteristics in terms of gains and implement a choice interface that mimics the CBS technology.\footnote{One could argue that in the real world no `lottery' guarantees only gains. However, we observe that presenting a financial product in a certain way can produce an illusion whereby the decision-maker thinks that they are in the presence of investment alternatives with `almost certain' returns.} On the other hand, a regulator would be able to limit the choice of risky mortgages by requiring - in this case - the relevant online choice architects to implement ABS technologies. Evidently, these speculations rest on the assumption that our results are externally valid. Therefore, a critical follow-up question is to investigate the robustness of the derived results outside the laboratory by testing our hypotheses in the field.

The rest of the paper is organised as follows. Section 2 reviews the related literature; Sections 3 and 4 present experiments 1 and 2, respectively; Section 5 offers an interpretation of the results; Section 6 concludes. The supplementary material (available upon request) contains the instructions, further details pertaining to the experimental design, further robustness checks, and three short videos aimed at illustrating the information-search technologies.

\section{Related Literature}

This paper relates to multiple strands of literature. First, experimental economists have investigated how people make decisions in choice problems involving multi-attribute alternatives. For instance, \citet{GabaixLaibsonMolocheWeinberg2006} use the Mouse-Tracing paradigm to experimentally investigate how experimental subjects acquire information. \citet{Sanjurjo2017} utilises \citet{GabaixLaibsonMolocheWeinberg2006}'s data to examine whether subjects search optimally, and find that subjects search too much within an alternative and fail to optimally switch to the next alternative to be investigated. On the other hand, \citet{ArieliBenAmiRustein2011} employ eye-tracking to analyse whether the subjects' eye movement is consistent with either an ABS or a CBS model when they choose between binary lotteries. They find that when the calculation of the expected value is relatively difficult (resp., simple) subjects' eye movement is consistent with a CBS (resp., a mixture of ABS and CBS) procedure. Unlike these studies, this paper adopts a different approach by exploring the effects of {\itshape inducing} individuals to use certain procedures.

As far as the latter approach is concerned, there are two related studies \citep{ReeckWallJohnson2017,MittonePapi2020}. On the one hand, \citet{ReeckWallJohnson2017} conduct an an experiment in which subjects are induced to use integrative vs comparative search strategies when making intertemporal choices. They find that encouraging subjects to search in different ways affects their choices over time. Unlike their work, this study examines risky choices. On the other hand, \citet{MittonePapi2020} find that inducing subjects to use CBS procedures (as opposed to ABS procedures) in lottery choice problems makes them to choose riskier lotteries. Their main focus is to study how the exposure to different heuristics relates with the complexity of the choice problem (measured in terms of {\itshape attribute-based} and {\itshape alternative-based} complexity) and, importantly, their design involves lotteries whose outcomes are gains only. In contrast, this study considers mixed lotteries and explores the interplay between the heuristic subjects are induced to use and the lottery-outcome domain, while holding complexity constant.

Third, a number of studies have investigated specific aspects that are relevant to digital nudging.\footnote{See also \citet{Waldman2020}, who discusses how cognitive biases are exploited by online platforms to induce individuals to disclose information.} For instance, \citet{SamekHurKimYi2016} examines the effects of different sorting technologies on subjects' choices by using the `choice-process' method, whereby not only final but also provisional choices are recorded. There are three main differences with this study. First, \citet{SamekHurKimYi2016}'s objects of choice are not lotteries, but vectors of numbers. Second, the interactive technologies used in \citet{SamekHurKimYi2016} are different from those used in this paper. Third, unlike \citet{SamekHurKimYi2016}, this paper does not utilise the `choice-process' method, but employs a variant of the Mouse-Tracing paradigm. On the other hand, motivated by the widespread use of pressure-selling techniques, \citet{KlimmKocherOpitzSchudy2023} explore the effects of perceived urgency and regret in a sequential search task. In contrast, our study investigates a different research question. The closest study to our work is \citet{AimoneBallKingCasas2016} who - following up on \citet{ArieliBenAmiRustein2011}'s findings - examine the effects of manipulating how information is acquired in a lottery-choice experiment. This study differs from ours in several respects: (i) the information-acquisition technologies are different from ours\footnote{For instance, unlike the CBS technology utilised in this paper, the `CCorder' technology employed in \citet{AimoneBallKingCasas2016} enables a subject to view {\itshape either} the outcome {\itshape or} the corresponding probability at any point in time.}; (ii) unlike \citet{AimoneBallKingCasas2016}, this study considers lotteries that have the same expected value and belong to the gain-loss lottery-outcome domain; and (iii) unlike in this study, in \citet{AimoneBallKingCasas2016} all subjects are shown - just before making a decision - all lottery attributes {\itshape at once}, regardless of the information-acquisition technology they are initially assigned to. Interestingly, although for the reasons outlined above their study is not comparable to ours, \citet{AimoneBallKingCasas2016} also find that exposing subjects to different information-acquisition formats has an effect on the outcome of their decisions.

Fourth, in the judgment and decision-making literature, an open question pertains to whether individuals are more likely to use holistic procedures in risky choices \citep{GloecknerHerbold2011}, or whether - instead - they tend to consider attributes separately \citep{BrandstaetterGigerenzerHertwig2006}. Although not conclusive, the results of this study seem to provide support in favour of the former view.

\section{Experiment 1}

As discussed above, in an earlier study \citep{MittonePapi2020}, we find that, when lottery outcomes comprise only gains, inducing subjects to use CBS procedures (as opposed to ABS procedures) makes them to choose {\itshape riskier} lotteries. Motivated by these findings, our first \href{https://osf.io/5xzqa}{\underline{pre-registered hypothesis}} is that these results extend to mixed lotteries, i.e., lotteries whose outcomes are gains and losses. The purpose of experiment 1 is to test this hypothesis.\footnote{The pre-registration contains an additional hypothesis (hypothesis 2) pertaining to the interplay between the heuristic subjects are induced to use and the classical preference-reversal phenomenon \citep{LichtensteinSlovic1971}. Such additional hypothesis and the resulting analysis will not be discussed here, but in a separate follow-up paper.}

\subsection{Experimental Design}

{\bfseries Subjects}. A total of 181 subjects were recruited from the Cognitive and Experimental Economics Laboratory (CEEL)'s subject pool of the University of Trento.\footnote{Most subjects were undergraduate students in economics. Further details about the subjects' characteristics can be found in the supplement.} A total of 64 subjects were assigned to the ABS treatment, 60 subjects to the CBS treatment, and 57 subjects to the Baseline treatment. Experiment 1 was between-subject, and took place online.\footnote{The relevant dates are 19th (CBS), 20th (ABS), and 21st (Baseline) of October 2021, 23rd (CBS), 24th (ABS), and 25th (Baseline) of November 2021, and 12th (ABS and CBS) and 13th (Baseline) of April 2022.} The software used in this experiment was designed by the authors of the paper and the CEEL manager Mr Marco Tecilla.


\vspace{0.5cm}


{\bfseries Lottery-Choice Task}. The experiment comprises: a lottery-choice task and a number of control tasks. While - as discussed below - {\itshape some} features of the lottery-choice task are treatment-specific, the control tasks are identical across treatments.

The lottery-choice task consists of a sequence of binary lottery choice problems. In all treatments of this experiment, a choice problem consists of two lotteries: a safe lottery $s$ and a risky lottery $r$. Every choice problem is presented in one screenshot, and subjects are asked to make a decision by choosing the lottery that they prefer the most within the time limit, which is 40 seconds. Table \ref{tab:Lotteries_Used} lists the safe and the risky lotteries that have been constructed for this experiment, which are common across treatments and whose outcomes are expressed in experimental points (1 point = 0.35 euros).\footnote{Observe that the safe lotteries can be interpreted as a $P$-bets and the risky lotteries as $\$$-bets \citep{LichtensteinSlovic1971}.} The lotteries have the properties that (i) they all have an expected value of 6 experimental points and (ii) every risky lottery is a {\itshape mean-preserving spread} of every safe lottery.\footnote{A key property of lotteries that can be ordered by the mean-preserving spread relation is as follows: a lottery $L^{\prime}$ is a mean-preserving spread of a lottery $L$ if and only if a risk-averse decision-maker prefers $L$ over $L^{\prime}$, irrespective of their degree of risk aversion \citep{RothschildStiglitz1970}.} 

\begin{table}
\begin{center}
\begin{tabular}{|c||c|c||c|c|}
\hline
$\#$ & Prob.& \textcolor{Green}{Gain} & Prob. & \textcolor{red}{Loss}\\ 
\hline
\hline
 & \multicolumn{4}{c|}{Safe Lotteries}\\
\hline 
$s_{1}$ & 0.80 & \textcolor{Green}{7.60} & 0.20 & \textcolor{red}{-0.40}\\
\hline 
$s_{2}$ & 0.70 & \textcolor{Green}{8.70} & 0.30 & \textcolor{red}{-0.30}\\
\hline 
$s_{3}$ & 0.60 & \textcolor{Green}{10.40} & 0.40 & \textcolor{red}{-0.60}\\
\hline
\hline
 & \multicolumn{4}{c|}{Risky Lotteries}\\
\hline 
$r_{1}$ & 0.40 & \textcolor{Green}{17.22} & 0.60 & \textcolor{red}{-1.48}\\
\hline 
$r_{2}$ & 0.30 & \textcolor{Green}{23.15} & 0.70 & \textcolor{red}{-1.35}\\
\hline 
$r_{3}$ & 0.20 & \textcolor{Green}{35.00} & 0.80 & \textcolor{red}{-1.25}\\
\hline
\end{tabular}
\end{center}
\begin{tablenotes}
      \centering
			\scriptsize
      \item \textcolor{Green}{Gains}/\textcolor{red}{losses} are expressed in exp.currency (1 point = 0.35 euros).
    \end{tablenotes}
\caption{Lotteries used in experiment 1}
\label{tab:Lotteries_Used}
\end{table}

We constructed a total of 9 binary lottery choice problems by matching each of the three safe lotteries with each of the three risky lotteries. Throughout, a choice problem will be denoted as an ordered pair $\left(s_{k},r_{l}\right)$, where $k,l \in \{1,2,3\}$, and $s_{k}$ and $r_{l}$ denote a safe and risky lottery, respectively, of table \ref{tab:Lotteries_Used}. In every treatment, the order according to which the choice problems is presented to subjects is random. Within each screenshot, the safe and risky lotteries are arranged side by side and their position (left or right) is randomised as well. In all treatments of this experiment, the outcome-probability pairs of each lottery are arranged on top of each other, with the property that - from the top to the bottom - the gain-probability pair is listed first and the loss-probability pair is listed second. At the end of the experiment, a lottery choice problem is selected at random, the lottery that the experimental subject chose at that particular choice problem is played, and the subject is paid accordingly. If subjects do not choose any lottery within the time limit at a particular choice problem, then a total of $1.5$ experimental points are deducted from the subject's earnings in the event that at the end of the experiment the choice problem under consideration is extracted for payment.

In order to induce subjects to use different procedures, we designed three treatments: ABS, CBS, and Baseline. These treatments are identical (and as described above), except for the technology that subjects use for exploring the lotteries available at each choice problem.

\vspace{0.3cm}

{\bfseries ABS Technology}. At every choice problem subjects are faced with in the lottery-choice task, information is initially hidden. In particular, subjects are faced with two buttons at the top of the screen corresponding to the two lotteries available at that particular choice problem. To discover the outcomes and the corresponding probabilities of a lottery, subjects click on the button corresponding to the lottery that they want to explore, and, at the bottom of the screen, the experimental software shows the outcomes and the corresponding probabilities of the selected lottery. Next to the discovered lottery, a `pie chart' whose slices graphically represent the probabilities of the different outcomes is also displayed. By clicking on the other button, the experimental software automatically {\itshape hides} the outcomes and corresponding probabilities of the lottery they initially clicked on, and {\itshape shows} the outcomes and corresponding probabilities of the lottery they have subsequently clicked on. Subjects are free to explore the lotteries as they like, even by discovering the same lottery more than once. The supplement contains a video (`ABS$\_$Treatment.mp4') demonstrating how subjects can explore information when they are administered the ABS technology. The ABS treatment is meant to induce an ABS procedure, as it encourages subjects to explore all the characteristics of one lottery, before exploring the next lottery.

\vspace{0.3cm}

{\bfseries CBS Technology}. Like in the ABS treatment, at every choice problem subjects are faced with in the lottery-choice task, information is initially hidden. Unlike in the ABS treatment, subjects explore information by discovering one outcome-probability pair at a time. In particular, by clicking on a gain-probability (resp., loss-probability) pair, the software automatically shows the gain-probability (resp., loss-probability) pairs of {\itshape both} lotteries available at a particular choice problem. Next to each lottery, a `pie chart' whose slices graphically represent the probability of the corresponding gain (resp., loss) is also displayed. By clicking on the other outcome-probability pair, the software automatically {\itshape hides} the outcome-probability pairs previously displayed and {\itshape shows} the outcome-probability pairs they have subsequently clicked on. Subjects are free to explore the outcome-probability pairs as they like, even by discovering the same outcome-probability pair more than once. The supplement contains a video (`CBS$\_$Treatment.mp4') demonstrating how subjects can explore information when they are administered the CBS technology. The CBS treatment is meant to induce a CBS procedure, as it encourages subjects to look up the value that an outcome-probability pair takes across both lotteries available, before exploring the next outcome-probability pair.

\vspace{0.3cm}

{\bfseries Baseline Technology}. Unlike in the ABS and CBS treatments, in the Baseline treatment the outcome-probability pairs of both lotteries available at a choice problem are immediately visible. As such, the Baseline treatment is not meant to induce any specific procedure, and will be used as a standard of comparison. The supplement contains a video (`Baseline$\_$Treatment.mp4') demonstrating how subjects can explore information when they are administered the Baseline technology.

We do not claim that assigning a subject to the ABS (resp., CBS) treatment forces them to use an ABS (resp., a CBS) procedure. Rather, we believe that the ABS technology makes the use of an ABS procedure more natural than the CBS technology. Likewise, the CBS technology makes the use of a CBS procedure more natural than the ABS technology.

\vspace{0.5cm}

{\bfseries Controls and Implementation }. In order to minimise the disruptions directly or indirectly caused by Covid (e.g. sudden imposition of restrictions that prevent the use of the laboratory), the experiment was conducted online by following a modification of \citet{ZhaoLopezvargasFriedmanGutierrez2020}'s protocol. In particular, the day before the experiment, the subjects were sent an email containing a link to a Zoom meeting and a link to the actual experiment that was operational only at the time in which the sessions were scheduled. On the day of the experiment, at the scheduled time subjects were asked to connect to the Zoom meeting with an experimentalist. They were instructed to keep their cameras on for the entire duration of the experiment. In this way the experimentalist was able to communicate with the subjects, and monitor their activities (e.g. check that no individual other than the experimental subject took part to the experiment). Subjects read the instructions, and then an experimentalist read them loudly. Before undertaking any task and after reading the instructions, subjects had to answer a number of control questions aimed at verifying whether they have understood the instructions.\footnote{The instructions as well as the control questions can be found in the supplement in both Italian (original language) and English.} They were allowed to proceed only if they answered all the questions correctly. 

Subjects began by undertaking the lottery-choice task described above. After completing the lottery-choice task, they were asked to undertake an extended cognitive reflection test (CRT7) developed by \citet{ToplakWestStanovich2014}.\footnote{As indicated in the first \href{https://osf.io/5xzqa}{\underline{pre-registration}}, after the lottery-choice task and before undertaking the CRT7, subjects were also asked to express their willingness to pay for the lotteries encountered in the lottery-choice task by using the \citet{BeckerDegrootMarschak1964}'s method (BDM). This task was included in the experimental design to test an additional hypothesis included in the pre-registration that relates the heuristic subjects are induced to use with the classical preference-reversal phenomenon \citep{LichtensteinSlovic1971}. Like for the other tasks, feedback was given at the end of the experiment. However, as discussed above, the combined analysis of the lottery-choice task and BDM will not be discussed here, but in a separate follow-up paper. In any case, for completeness, in the supplement we include a detailed description of the BDM and corresponding instructions.} Subjects were rewarded a total of 0.07 points for every correct answer they provided to the CRT7 questions. Then, subjects went through the Bomb Risk Elicitation Task (BRET) developed by \citet{CrosettoFilippin2013}.\footnote{For a detailed description of the BRET, see the supplement.} After completing the BRET, subjects were administered an anonymous questionnaire aimed at recording their demographics. To have an indication of subjects' level of attention in the experiment, we included in the questionnaire a 7-point scale introspective attention question, where 1 (resp., 7) corresponded to minimal (resp., maximal) attention. Furthermore, we added an additional multiple-choice introspective salience question aimed at asking subjects whether the aspect that mattered the most to them in making lottery decisions was: (1) the outcomes, (2) the probabilities, (3) both the outcomes and the probabilities, or (4) neither the outcomes nor the probabilities of the lotteries. We then asked subjects to guess what was the most popular answer to the introspective salience question by the other experimental subjects in the same session by rewarding a correct guess with a total of 3 experimental points.\footnote{Due to an unintended mistake in the experimental software, the introspective salience question and the corresponding guessing question were recorded in the second and third sessions of experiment 1 (those taking place in November 2021 and April 2022), but not in the first session (that taking place in October 2021). As a result, the number of observations for these two questions in experiment 1 are: 37, 41, and 36 in the ABS, CBS, and Baseline, respectively.} 

Subjects were given feedback at the end of the experiment and were informed about their total monetary payoff, inclusive of the show-up fee (10 experimental points), and were paid accordingly via a bank transfer. Subjects earned on average 9.07, 9.14, and 8.63 euros in the ABS, CBS, and Baseline treatment, respectively.

\begin{figure}[htb]
	\centering
		\includegraphics[width=150mm]{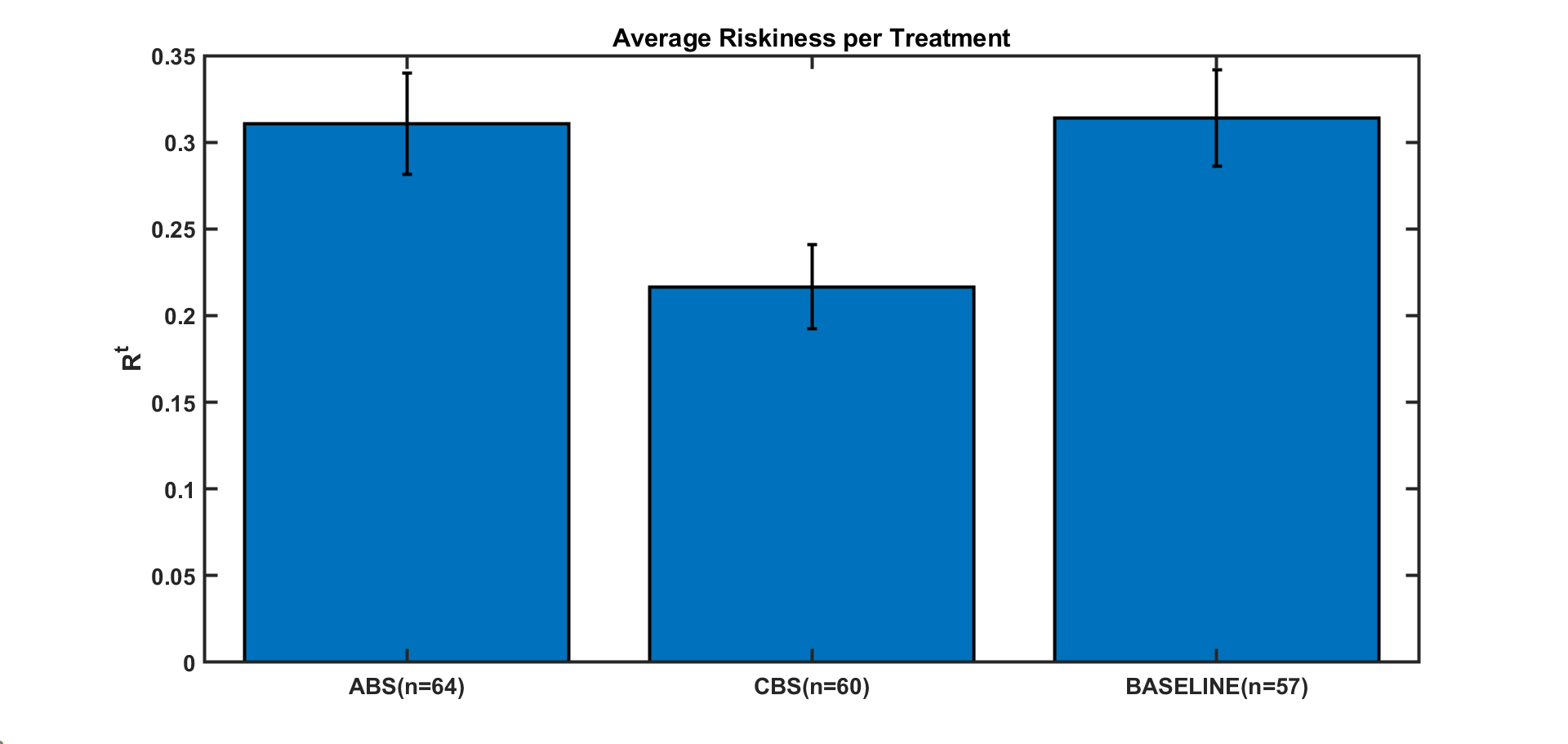}
		\caption{Average Riskiness $R^{t}$ with standard error bars; Exp.1}
		\label{Fig:Diagram_Full}
		\end{figure}

\subsection{Results}

{\bfseries Choice Analysis}. Throughout, let $\mathcal{N}_{t}$ denote the set of experimental subjects assigned to treatment $t$, where $t \in \{ABS, CBS, Base\}$. Recall that a choice problem is denoted by an ordered pair $(s_{k},r_{l})$, where $k,l \in \{1,2,3\}$, and $s_{k}$ and $r_{l}$ denote a safe and risky lottery, respectively, of table \ref{tab:Lotteries_Used}. We being by investigating the extent to which subjects made risky choices in the lottery-choice task. To do so, for every subject $i$ facing a choice problem $(s_{k},r_{l})$ in treatment $t$, we construct a {\itshape riskiness index} $R_{i}^{t}(s_{k},r_{l})$ as follows.

\begin{equation}
   R_{i}^{t}(s_{k},r_{l}):= \left\{
     \begin{array}{ll}
       0 & \mbox{, if subject $i$ chose $s_{k}$ at problem $(s_{k},r_{l})$}\\
       1 & \mbox{, otherwise.}
     \end{array}
   \right.
	\label{eq:riskiness_index}
\end{equation} 

That is, the riskiness index $R_{i}^{t}(s_{k},r_{l})$ is equal to $0$ (resp., $1$) if subject $i$ in treatment $t$ chose the safe (resp., risky) lottery. Recall that in our design every risky lottery is a mean-preserving spread of every safe lottery. As such, the riskiness index satisfies the property that it is increasing in the riskiness of the chosen lottery. 

In addition, define $R_{i}^{t}:=\frac{1}{9}\sum_{k} \sum_{l} R_{i}^{t}(s_{k},r_{l})$ as subject $i$'s average riskiness index across choice problems in treatment $t$, and $R^{t}:=\frac{1}{|\mathcal{N}_{t}|}\sum_{i \in \mathcal{N}_{t}} R_{i}^{t}$ as the average riskiness index in treatment $t$. Figure \ref{Fig:Diagram_Full} displays the average riskiness index $R^{t}$ in each treatment $t$ with standard error bars.

\begin{figure}[htb]
	\centering
		\includegraphics[width=150mm]{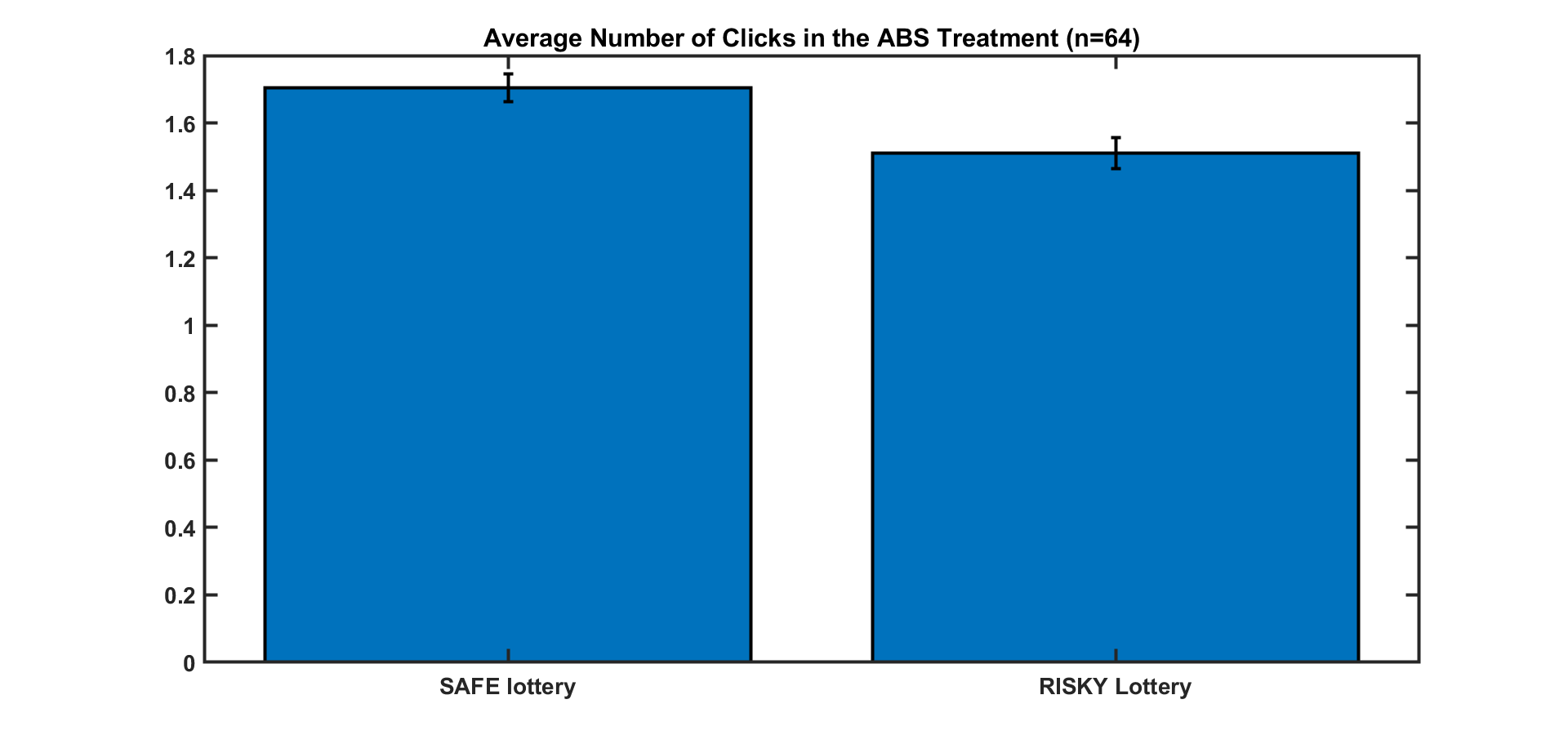}
		\caption{Average Number of Clicks in the ABS Treatment; Exp.1}
		\label{Fig:ABS_Lookups}
		\end{figure}

By looking at figure \ref{Fig:Diagram_Full}, three observations can be drawn. First, the average riskiness index $R^{t}$ is in absolute value significantly smaller than $0.50$ in all treatments. Given that in this experiment every risky lottery is a mean-preserving spread of every safe lottery, subjects' choices reveal that their propensity towards risk tends to be on average of aversion to risk. Second, the subjects assigned to the ABS and Baseline treatment chose in a similar way in terms of riskiness. Third - and most importantly - the subjects assigned to the CBS treatment systematically chose in a {\itshape safer} way compared to the subjects assigned to the ABS and Baseline treatment. We run a Mann-Whitney test for independent samples and find that the difference in average riskiness between ABS-CBS and Baseline-CBS is statistically significant at conventional levels (standard test statistics are $2.383^{**}$ and $2.601^{***}$, respectively). Robustness checks consisting of a bootstrap analysis for independent samples t-test confirm these results (estimated p-values are $0.015^{**}$ and $0.009^{***}$, respectively).\footnote{In conducting each bootstrap procedure for this experiment, a total of 9,999 bootstrap samples were generated \citep{Moffatt2016}.}

\begin{figure}[htb]
	\centering
		\includegraphics[width=150mm]{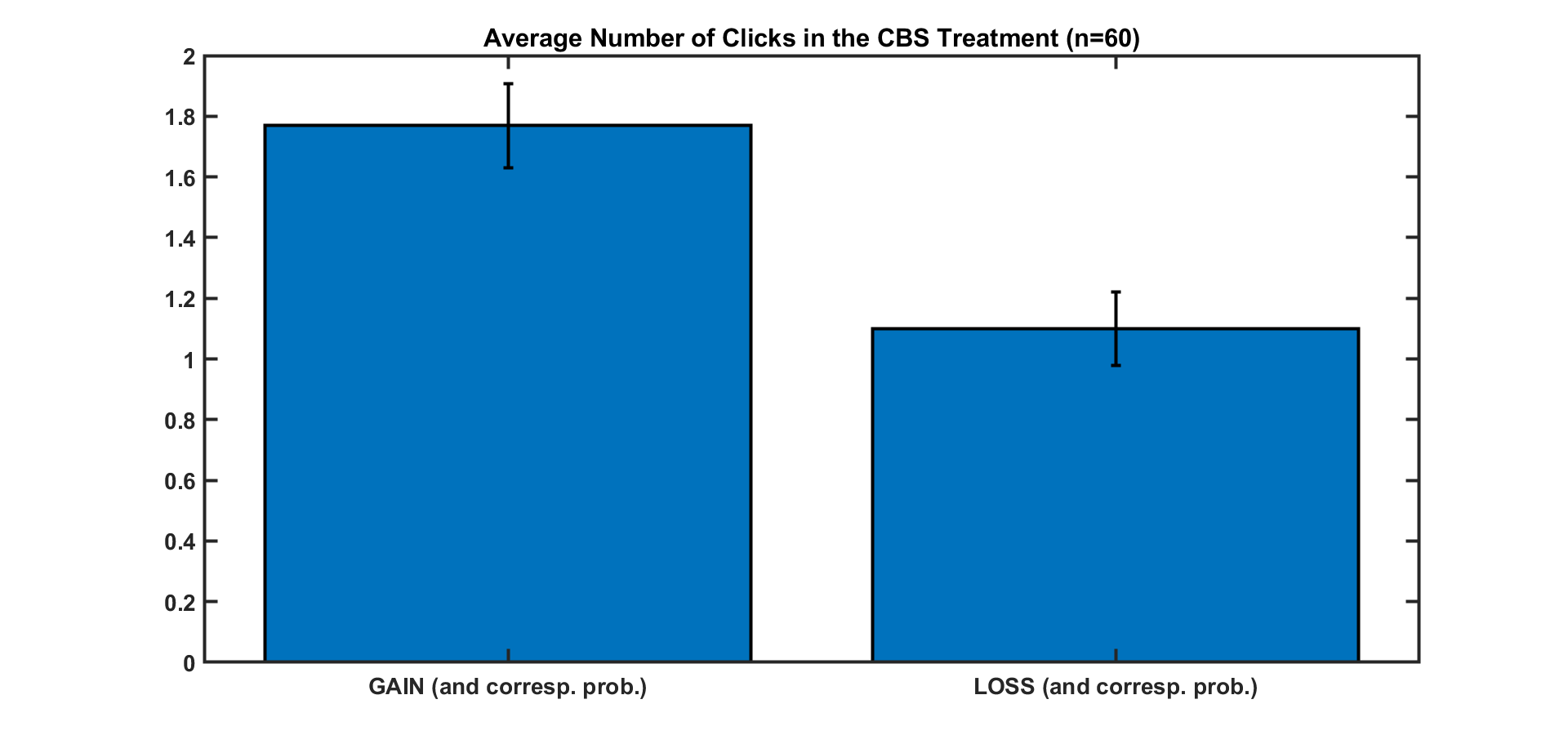}
		\caption{Average Number of Clicks in the CBS Treatment; Exp.1}
		\label{Fig:CBS_Lookups}
		\end{figure}

\vspace{0.5cm}


{\bfseries Information-Search Analysis}. We begin the information search analysis by investigating the average number of times subjects looked up the safe and risky lottery in the ABS treatment. As illustrated in figure \ref{Fig:ABS_Lookups}, subjects clicked on average 1.70 times on the safe and 1.51 times on the risky lottery. Although the averages are not too far apart, a Wilcoxon signed-rank test for dependent samples confirms that the number of clicks on the safe and risky lottery are significantly different at conventional levels (test stat.= $4.635^{***}$).


\begin{table}[htb]
\small
\begin{tabular}{lcccc}
\midrule
\multicolumn{5}{l}{Panel Probit Regression on the riskiness index in the ABS treatment (Exp.1)}\\
\toprule
{\itshape Riskiness index}			& Model 1				& Model 2				& Model 3				& Model 4\\
\toprule
{\itshape $\#$safe lott.clicks}	&	-2.596471$^{***}$	& -2.604038$^{***}$	& -3.154586$^{***}$	& -3.109959$^{***}$\\
																& (.313102)			& (.3360061)		& (.4571613)		& (.4478019)\\[3ex]
{\itshape $\#$risky lott.clicks}&	2.806335$^{***}$		& 2.695346$^{***}$		& 3.011842$^{***}$		& 2.957496$^{***}$\\
																& (.3149929)    & (.3305261)		& (.4295831)		& (.4202122)\\[3ex]
{\itshape Clicked on both lott.}&	-.0745059			& .4866022			& .70863				& .8485965\\
																& (1.900412)		& (3.937476)		& (5.477841)		& (6.076832)\\[3ex]
{\itshape Time on safe lott.($s$)}& 						& -.0201542			& .0182827		  & .0146173\\
																&  							& (.0243167)		& (.0280446)		& (.028186)\\[3ex]
{\itshape Time on risky lott.($s$)}& 						& .0522189$^{*}$			& .0944618$^{***}$  	& .0910314$^{***}$\\
																&  							& (.0270158)		& (.0320183)		& (.0317932)\\[3ex]
{\itshape Controls}							& No						& No						& Yes 					& Yes\\[3ex]
{\itshape Demographics}					& No						& No						& No 						& Yes\\[3ex]
{\itshape Constant}							&	-.7806593			& -1.384203			&  -4.684561		& -1.892109\\
																& (1.89635)		  & (3.937924 )		& (5.558398)		& (6.254942)\\
\midrule
Log likelihood									& -180.81086		&  -178.75117		& -159.47795 		& -156.69288\\
Akaike inf.crit.								&	 371.6217			& 	371.5023  	& 370.9559 			& 369.3858 \\
Observations										& 576						& 576						& 576						& 576\\
\toprule
\end{tabular}
\begin{tablenotes}
			\scriptsize
      \item Standard errors in parentheses, $^{*}$ $p<0.10$, $^{**}$ $p<0.05$, and $^{***}$ $p<0.01$. Controls: period at which choice was made, choice problem, number of collected boxes (BRET), 7-item CRT score, and player declared attention (7-point scale, where 7 means `maximum attention'). Demographics: age and gender. Lottery outcomes are gains and losses.
    \end{tablenotes}
\caption{Choice and Information-Search in the ABS Treatment($n=64$); Exp.1}
\label{tab:ABS_xtProbit}
\end{table}
  

We next examine the average number of times subjects explored the gain-probability pair and the loss-probability pair in the CBS treatment. As shown in figure \ref{Fig:CBS_Lookups}, subjects clicked on the gain-probability pair an average of 1.77 times and on the loss-probability pair an average of 1.10 times. These results provide clear evidence that subject systematically explored more often the gain-probability pair than the loss-probability pair (Wilcoxon signed-rank test stat.= $6.445^{***}$).

\vspace{0.5cm}

{\bfseries Information-Search and Choice Analysis}. We next combine the information-search and choice analysis, in order to shed light on how subjects’ clicking behaviour is related with their choices in the ABS and CBS treatments. We do so by running a set of random-effects panel probit regressions. In particular, as far as the ABS treatment is concerned, the riskiness index $R_{i}^{ABS}(s_{k},r_{l})$ (equation \ref{eq:riskiness_index}) has been regressed on the number of times subjects have looked up the safe and risky lottery, a dummy variable capturing whether or not they have explored both lotteries, the amount of time (in seconds) they have spent on looking up the safe and risky lottery, and additional control variables. Results are displayed in table \ref{tab:ABS_xtProbit}. As one expects, the number of times subjects have clicked on the safe (resp., risky) lottery is negatively (resp., positively) related to the probability that the risky lottery is chosen. The corresponding average marginal effects are -.4279682 ($z = -9.16^{***}$) and .4069875 ($z = 9.29^{***}$), respectively. The only other regressor that has a statistically significant coefficient is the time spent on the risky lottery, although the corresponding marginal effect is relatively small ($<0.02$).


\begin{table}[htb]
\small
\begin{tabular}{lcccc}
\midrule
\multicolumn{5}{l}{Panel Probit Regression on the riskiness index in the CBS treatment (Exp.1)}\\
\toprule
{\itshape Riskiness index}			& Model 1				& Model 2				& Model 3				& Model 4\\
\toprule
{\itshape $\#$ gain clicks}			&	.0087717			& -.0387183			& -.0312628		& -.0306694\\
																& (.0960892)		& (.122487)			& (.1343821)  & (.1341742)\\[3ex]
{\itshape $\#$ loss clicks}			&	-.0263597			& -.0893256			& -.1675496		& -.1637419\\
																& (.1074184)    & (.1430135)		& (.1592626)  & (.1596865)\\[3ex]
{\itshape Clicked on both} 			&	.4041006$^{*}$			& .3758542$^{*}$			& .3889483		& .3985032$^{*}$\\
																& (.210637)			& (.2142926)		& (.2377372)	& (.2395036)\\[3ex]
{\itshape Time on gains($s$)} 	&  							& .0276693$^{*}$			& .0647531$^{***}$	& .064431$^{***}$\\
																&  							& (.0158024)		& (.0189068)	& (.0189098)\\[3ex]
{\itshape Time on losses($s$)}	&  							& .0260843			& .0486575$^{**}$	& .0480832$^{**}$\\
																&  							& (.0189092)		& (.0215103)	& (.0215431)\\[3ex]
			
{\itshape Controls}							& No						& No						& Yes 					& Yes\\[3ex]
{\itshape Demographics}					& No						& No						& No						& Yes\\[3ex]
{\itshape Constant}							&	-1.089952$^{***}$	& -1.238329$^{***}$	& -2.653098$^{***}$	& -2.299955\\
																& (.1521088)		& (.1681951)		& (.7629245)		& (1.465449)\\
\midrule
Log likelihood									& -270.88113		& -268.06494		& -242.04729		& -241.95995\\
Akaike inf.crit.								&	551.7623			& 550.1299			& 536.0946			&  539.9199\\
Observations										& 540						& 540						& 540						& 540\\
\toprule
\end{tabular}
\begin{tablenotes}
			\scriptsize
      \item Standard errors in parentheses, $^{*}$ $p<0.10$, $^{**}$ $p<0.05$, and $^{***}$ $p<0.01$. Controls: period at which choice was made, choice problem, number of collected boxes (BRET), 7-item CRT score, and player declared attention (7-point scale, where 7 means `maximum attention'). Demographics: age and gender. Lottery outcomes are gains and losses.
    \end{tablenotes}
\caption{Choice and Information-Search in the CBS Treatment ($n=60$); Exp.1}
\label{tab:CBS_xtProbit}
\end{table}


A similar random-effects panel probit regression has been performed for the CBS treatment. In particular, the riskiness index $R_{i}^{CBS}(s_{k},r_{l})$ has been regressed on the number of times subjects have looked up the gain and the loss (and corresponding probabilities), a dummy variable capturing whether or not they have explored both outcomes and corresponding probabilities, the amount of time (in seconds) they have spent on the gain and the loss (and corresponding probabilities), and additional control variables. Results are displayed in table \ref{tab:CBS_xtProbit}.

Unlike one would expect, the number of times subjects clicked on the gains or losses (and corresponding probabilities) does not seem to be related to the probability of choosing the risky lottery. To the contrary, the extent to which subjects explore both the gain and the loss (and corresponding probabilities) appears to be a significant factor, in the sense that when subjects happen to explore {\itshape both} outcomes (and corresponding probabilities), they are {\itshape more} likely to choose the risky lottery in the CBS treatment. An analysis of the marginal effects reveals that doing so {\itshape increases} on average the probability of choosing the risky lottery by 8.6 percentage points ($z=1.64^{*}$). Consistently with the latter, the time spent on both gains and losses (and corresponding probabilities) is also positively related with the probability of choosing the risky lottery.

\vspace{0.5cm}

{\bfseries Controls}. Evidence indicates that female subjects tend to be more risk-averse than male subjects \citep{CrosonGneezy2009}. For this reason, we made sure that all our treatments are balanced in terms of gender.\footnote{Overall, we recruited 96 female and 85 male subjects by allocating them as follows: 35 and 29 in the ABS, 33 and 27 in the CBS, and 28 and 29 in the Baseline, female and male subjects, respectively.} Another potential issue is that our results could be driven by systematic differences in risk attitudes across treatments. In the supplement, we demonstrate that in our experiment there is no difference in the distribution of subjects' risk preferences elicited via BRET across treatments, and conduct additional robustness checks. 


\subsection{Summary of the Results}


The results of experiment 1 indicate that, contrary to our initial hypothesis, when lottery outcomes comprise gains and losses, inducing subjects to use a CBS procedure makes them choose {\itshape safer} options. Specifically, the information-search analysis indicates that in the ABS treatment subjects tend to look up the safe lottery a little more often than the risky lottery. By combining the information-search and choice analysis, we find that the probability that the risky lottery is chosen in the ABS treatment is positively (resp., negatively) related with the number of times subjects click on the risky (resp., safe) lottery. On the other hand, in the CBS treatment subjects systematically look up the gains (and corresponding probabilities) more often than the losses. Interestingly, although the subjects assigned to the CBS treatment choose on average in a safer way, we find that whenever they happen to look up {\itshape both} outcomes, the probability that they choose the risky lottery increases, which results in their behaviour being more in line with that of the subjects assigned to the other treatments.


\section{Experiment 2}

After conducting experiment 1 and analysing the resulting data, we concluded that our first hypothesis is falsified. The first hypothesis states that inducing subjects to use a CBS procedure makes them choose {\itshape riskier} lotteries. In contrast, unlike we expected, we found that in experiment 1 inducing subjects to use a CBS procedure made them choose {\itshape safer} lotteries. These unexpected results have led us to formulate an alternative explanation: it is the presence of losses  amongst the lottery outcomes that produces a discrepancy between the first hypothesis and the derived results. This conjecture is supported by the fact that it is well-known that individuals treat gains and losses differently in various choice settings ranging from a choice-from-description \citep{KahnemanTversky1979} to a choice-from-experience setup \citep{WulffMergenthalerCansecoHertwig2018}.\footnote{See also \citet{BatemanDentPetersSlovicStarmer2007} for a study that investigates the effects of introducing small losses.}

To test this explanation, we \href{https://osf.io/3xeav}{\underline{pre-registered a second experiment}} that is identical to experiment 1, except that to all outcomes of the lotteries used in experiment 1, the same positive constant is added, in order to turn the losses into (small) gains. This transformation produces lotteries that, from the point of view of a standard expected  utility  maximiser, are ordered in the same way as the original ones. We hypothesize that running a second experiment using the transformed lotteries reverses the results found in experiment 1. That is, consistently with \citet{MittonePapi2020}, we expect that inducing subjects to use CBS procedures makes them choose {\itshape riskier} lotteries.\footnote{Like the first pre-registration, also the second pre-registration contains a further hypothesis pertaining to the interplay between the heuristic subjects are induced to use and the classical preference-reversal phenomenon. As mentioned above, this additional hypothesis and the resulting analysis will not be discussed here, but in a separate follow-up paper.} The purpose of experiment 2 is to test this second hypothesis.


\subsection{Experimental Design}

{\bfseries Subjects}. A total of 129 subjects were recruited from the CEEL's subject pool of the University of Trento.\footnote{Most subjects were undergraduate students in economics.} A total of 42 subjects were assigned to the ABS treatment, 42 subjects to the CBS treatment, and 45 subjects to the Baseline treatment. Like experiment 1, experiment 2 was between-subject and took place online.\footnote{The relevant dates are 9th (ABS and CBS) and 10th (Baseline) of November 2022, and 7th (ABS and CBS) and 8th (Baseline) of March 2023.}

\vspace{0.5cm}

\begin{figure}[htb]
	\centering
		\includegraphics[width=150mm]{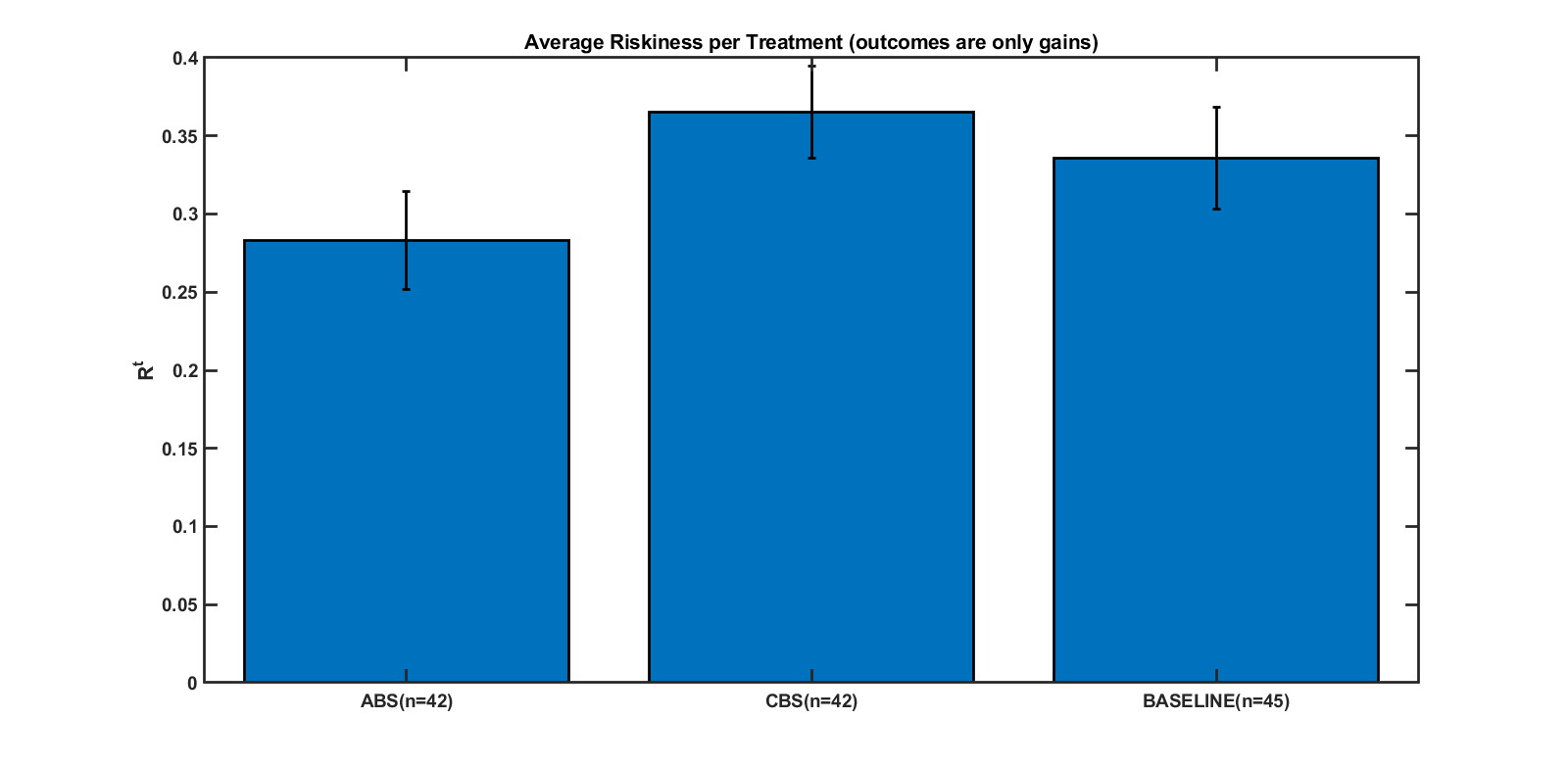}
		\caption{Average Riskiness $R^{t}$ with standard error bars; experiment 2}
		\label{Fig:Diagram_Full_2}
		\end{figure}

{\bfseries Design}. The design of experiment 2 is identical to that of experiment 1, except for the lotteries used. In particular, the lotteries used in experiment 2 have been obtained by adding the same positive constant (i.e., 2.00 experimental points) to all outcomes of the lotteries used in the experiment 1, while holding the probabilities constant.\footnote{The resulting transformed lotteries can be found in the supplement. Observe that, in order to make sure that subjects would earn approximately the same amount of money in the two experiments, the exchange rate of experiment 2 has been decreased to `1 exp. currency = 0.30 euros'.} In this way, all outcomes of the resulting lotteries are gains, and the resulting lotteries continue to have the properties that (i) every lottery has the same expected value, (ii) every transformed risky lottery is a mean-preserving spread of every transformed safe lottery. Furthermore, letting $s_{k}$ and $r_{l}$ denote a safe and a risky lottery used in experiment 1 and letting $\tilde{s}_{k}$ and $\tilde{r}_{l}$ denote the corresponding transformed lotteries used in experiment 2, a risk-averse decision-maker prefers lottery $s_{k}$ over lottery $r_{l}$ in experiment 1 if and only if a risky-averse decision-maker prefers transformed lottery $\tilde{s}_{k}$ over transformed lottery $\tilde{r}_{l}$ in experiment 2.

In experiment 2, subjects earned on average 8.87, 8.97, and 9.02 euros in the ABS, CBS, and Baseline treatment, respectively.

\begin{figure}[htb]
	\centering
		\includegraphics[width=150mm]{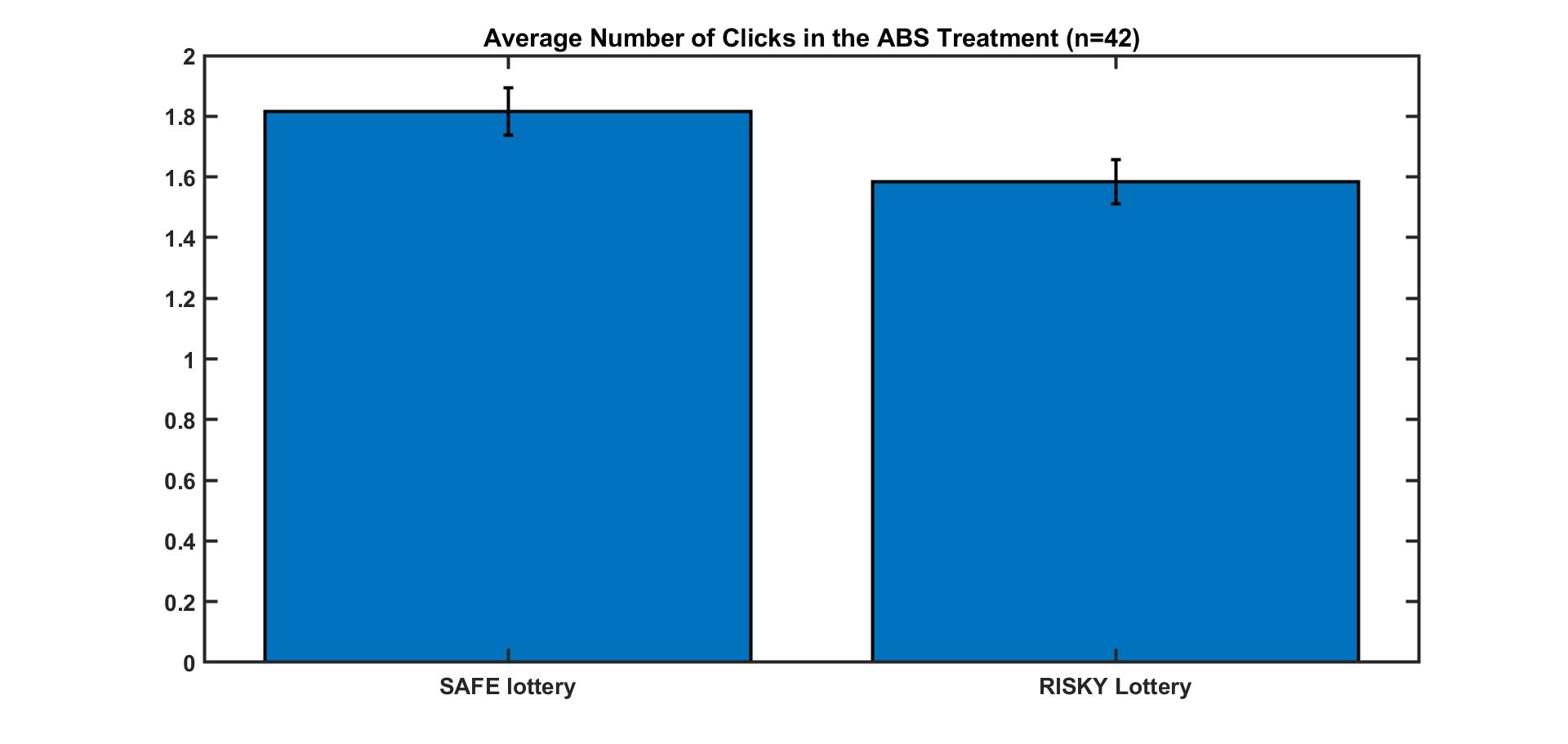}
		\caption{Average Number of Lookups in the ABS Treatment; experiment 2}
		\label{Fig:ABS_Lookups_2}
		\end{figure}

\subsection{Results}

{\bfseries Choice Analysis}. Figure \ref{Fig:Diagram_Full_2} displays the average riskiness index $R^{t}$ along with standard error bars. On the one hand, we find that - like in experiment 1 - average $R^{t}$ is smaller than 0.5 across treatments, revealing a tendency of subjects to be averse to risk. On the other hand, we find that that, in sharp contrast to the results of experiment 1, the subjects assigned to the CBS treatment choose {\itshape riskier} options compared to the subjects assigned to the ABS treatment. A Mann-Whitney test for independent samples and a bootstrap analysis for independent samples t-test confirm that the difference is statistically significant at conventional levels (the corresponding standard test statistics is $-1.815^{*}$ and the estimated p-value is $0.059^{*}$, respectively). As far as the CBS-Baseline comparison is concerned, although figure \ref{Fig:Diagram_Full_2} indicates that the average riskiness index of the Baseline treatment is smaller than that of the CBS treatment and greater than that of the ABS treatment, we do not detect any statistically significant difference. 

\begin{figure}[htb]
	\centering
		\includegraphics[width=150mm]{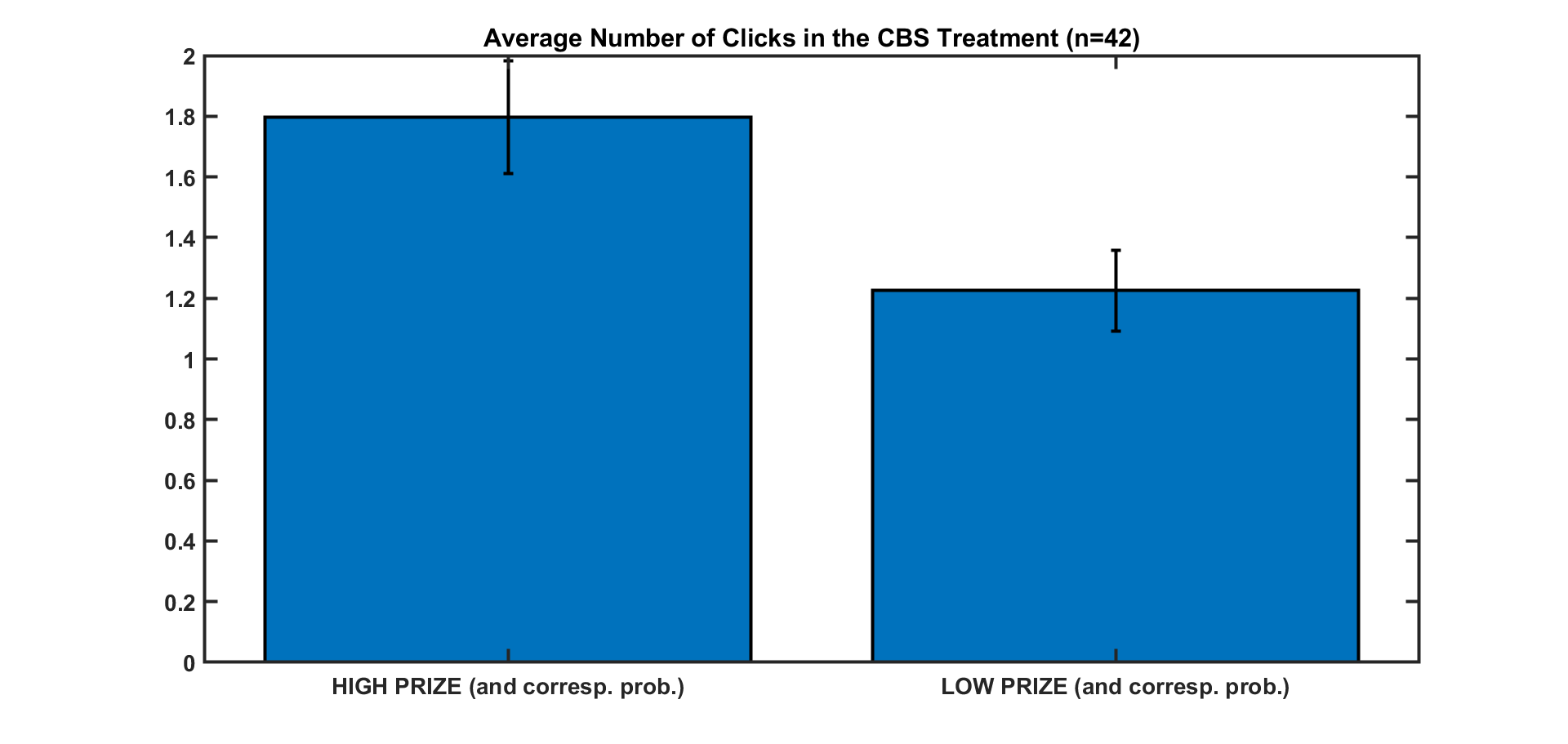}
		\caption{Average Number of Lookups in the CBS Treatment; experiment 2}
		\label{Fig:CBS_Lookups_2}
		\end{figure}

A natural follow-up question is to investigate whether there are systematic differences within treatments by transitioning from the gain-loss lottery-outcome domain (experiment 1) to the gain lottery-outcome domain (experiment 2). Interestingly, as far as the ABS treatment is concerned, there is no statistically significant difference in the average riskiness index $R^{t}$ between experiment 1 and experiment 2 (average $R^{t}$ is 0.31 and 0.28, respectively). The same pattern occurs in the Baseline treatment (the corresponding average $R^{t}$ is 0.31 and 0.34, respectively). In contrast, in the CBS treatment, experiment 1's average $R^{t}$ is markedly smaller than experiment 2's average $R^{t}$ (0.22 and 0.37, respectively). A Mann-Whitney test for independent samples and a bootstrap analysis for independent samples t-test confirm that this difference is statistically significant at conventional levels (the corresponding standard test statistics is $-3.807^{***}$ and the estimated p-value is $0.000^{***}$, respectively). 

\vspace{0.5cm}

{\bfseries Information-Search and Choice Analysis}. Figure \ref{Fig:ABS_Lookups_2} shows the average number of times subjects clicked on the safe and risky lottery in the ABS treatment (1.81 and 1.58 times, respectively). Figure \ref{Fig:CBS_Lookups_2}, on the other hand, displays the average number of times subjects looked up the high prize and the low prize and corresponding probabilities in the CBS treatment (1.80 and 1.22 times, respectively). Results are analogous to those of experiment 1: (i) subjects tend to click on the safe lottery a little more often than on risky lottery and (ii) subjects systematically look up more often the high prize than the low prize (and corresponding probabilities) in the CBS treatment. The differences in the number of clicks are statistically significant at conventional levels (the corresponding Wilcoxon signed-rank test statistics are $4.035^{***}$ and $5.054^{***}$, respectively).

Just like we did for experiment 1, we run a set of random-effects panel probit regressions - whose details can be found in the supplement - in order to gain further insights into how subjects’ information-search behaviour is related with their choices in the ABS and CBS treatments. As far as the ABS treatment is concerned, we similarly found that the number of times subjects have looked up the safe (resp., risky) lottery is negatively (resp., positively) related to the probability that the risky lottery is chosen. The corresponding average marginal effects are -.3079823 ($z =-11.42^{***}$) and .3684556 ($z =17.01^{***}$), respectively. As far as the CBS treatment is concerned, like in experiment 1, the extent to which subjects explore both outcomes (and corresponding probabilities) tends to be a significant factor. However, in sharp contrast to the results of experiment 1, when subjects happen to explore both outcomes (and corresponding probabilities), they are {\itshape less} likely to choose the risky lottery. In particular, an analysis of the corresponding average marginal effect indicates that exploring both outcomes and corresponding probabilities {\itshape decrease} the chances of choosing the risky lottery by about $18\%$ ($z = -2.45^{**}$).

Just like we did for experiment 1, we made sure that all treatments of experiment 2 are balanced in terms of gender, and conducted additional robustness checks, which can be found in the supplement.

\subsection{Summary of the Results}

In line with our second hypothesis, we find that, when lottery outcomes comprise only gains, the main result of experiment 1 is reversed, i.e., inducing subjects to use a CBS procedure makes them choose {\itshape riskier} options. In the ABS treatment we find that - like in experiment 1 - (i) subjects tend to look up the safe lottery a little more often than the risky lottery and (ii) the probability that the risky lottery is chosen in the ABS treatment is positively (resp., negatively) related with the number of times subjects click on the risky (resp., safe) lottery. Likewise, in the CBS treatment, our results indicate that subjects systematically look up the high prize (and corresponding probabilities) more often than the low prize. However, although the subjects assigned to the CBS treatment choose on average in a riskier way, we find that whenever they happen to look up {\itshape both} outcomes, the probability that they choose the risky lottery decreases, which results in their behaviour being more in line with that of the subjects assigned to the other treatments. Interestingly, while in experiment 1 the probability that the risky lottery is chosen {\itshape increases} when both outcomes are looked up, in experiment 2 the probability that the risky lottery is chosen {\itshape decreases} when both outcomes are looked up.

\section{Discussion}

By jointly considering experiment 1 and experiment 2, we find that subjects' choice behaviour in the ABS and Baseline treatment is not influenced by the lottery-outcome domain. In fact - as demonstrated above - by transitioning from the gain-loss lottery-outcome domain (experiment 1) to the gain lottery-outcome domain (experiment 2), we do not detect any statistically significant difference in average risky choice within these two treatments. In contrast, we find that in the CBS treatment the lottery-outcome domain {\itshape does} have an impact on subjects' choices. In particular, on the one hand, when the lottery-outcome domain comprises gains and losses (experiment 1), inducing subjects to use a CBS procedure makes them choose {\itshape safer} lotteries. On the other hand, when the lottery-outcome domain comprises gains only (experiment 2), inducing subjects to use a CBS procedure makes them choose {\itshape riskier} lotteries. Why do subjects exhibit this switch in behaviour when they are induced to use a CBS procedure? And why do they not exhibit any switch in the other two conditions?

These findings were unexpected and turned out to be more interesting than we initially conjectured. We indeed obtain preference reversals by simply exposing subjects to certain information-search technologies, while holding both the outcomes and the probabilities of the lotteries available constant. In this section we propose an {\itshape ex-post} rationalisation of our results that is based on prospect theory. Specifically, we argue that, at the core of the matter, is the combined effect on the formation of the reference point of, on the one hand, the heuristic subjects are induced to use and, on the other hand, the lottery-outcome domain. 

In prospect theory decision-makers evaluate the lottery outcomes with respect to a reference point, which defines the domains of gain and losses, with the property that in the gain (resp., loss) domain subjects are risk-averse (resp., risk-lover) \citep{KahnemanTversky1979}. A key aspect of the `coding phase' is therefore that of the formation of the reference point. Although in the early formulations of prospect theory the reference point is exogenously determined, in many cases there is a natural candidate for the reference point, such as the status-quo option and an individual's current asset position. In other scenarios, the reference point may be given by a decision maker's expectation or aspiration \citep{ODonoghueSprenger2018}.\footnote{See \citet{KoszegiRabin2006} as a prominent study aimed at endogenising the reference point by assuming it to coincide with the decision-maker's expectations.} As such, it is worth bearing in mind that the reference-point position is subjective and cannot be mechanically traced back to an intrinsic objective nature. 

Turning to our experiments, it is reasonable to assume that, when lottery outcomes involve gains and losses (experiment 1), subjects consider any positive amount as a gain and any negative amount as a loss, by implicitly setting the reference point at zero. On the other hand, when lottery outcomes involve gains only (experiment 2), it is plausible to expect that the reference point is given by some function of the lottery attributes, that has the property that it is nondecreasing in the lottery outcomes. In the latter case, examples of the candidate reference point include the average lottery outcome, the maxmin payoff, and the expected value.\footnote{\citet{BaillonBleichrodtSpinu2020}, for instance, investigate which reference point, such as expected value,  maxmin payoff, and most likely outcome, subjects use in a lottery-choice experiment in which the lottery-outcome domain involves gains only. Amongst others, see also \citet{HackBieberstein2015} who provide an in-depth look at reference point formation mainly by investigating the expectations' role. Specifically, they suggest an `integrated mechanism' whereby people's expectations are based on the average outcomes rather than any single potential outcome.} Therefore, by transition from a lottery-outcome domain consisting of gains and losses (experiment 1) to a lottery-outcome domain involving only gains (experiment 2), the reference point is likely to shift to the right by taking on a value that is strictly greater than zero.

The key question then becomes: why should a rightward shift of the reference point cause the subjects that are exposed to the CBS heuristic to exhibit a preference reversal? And why does this phenomenon not occur to the subjects that are assigned to the other conditions? Although only speculative, one possible explanation of this preference reversal is that the CBS treatment - which forces subjects to view an outcome at a time - makes the `{\itshape high-prize}' attribute relatively more salient. As evidenced by the information-search analysis, we indeed observe that subjects systematically click more often on the high prize (and corresponding probability) than on the low prize.\footnote{An analysis of subjects' modal response to the introspective salience question indicates that - in both experiment 1 and experiment 2 - the aspect that mattered the most to them in making lottery choices was `{\itshape both the outcomes and the probabilities}' for the subjects assigned to the ABS and Baseline treatment and `{\itshape the probabilities}' for the subjects assigned to the CBS treatment (details can be found in the supplement). Retrospective protocols do not appear to be a valid instrument for inferring the underlying cognitive processes \citep{RussoJohnsonStephens1989}. As such, subjects' responses to the introspective salience question should be used with caution, and are not incompatible with the proposed interpretation of the results.} Assuming, as we have just done, that the reference point is taken to be a non-decreasing function of the lottery outcomes, it follows that focusing on the high-prize attribute (by paying relatively little attention to the associated probability of occurrence and the other lottery attributes) translates into a further right-ward shift of the reference point.\footnote{According to the interpretation proposed in \citet{MittonePapi2020}, it is also assumed that - to the subjects that are exposed to the CBS heuristic - the high prize becomes relatively more salient.} This is equivalent to saying that the domain of losses prevails over the gain domain, which ultimately results in subjects exhibiting relatively more risky choice behaviour in experiment 2.

On the other hand, the subjects exposed to the ABS heuristics and the subjects who left free to compare all lotteries simultaneously (Baseline treatment) do not exhibit such a change in behaviour. The reason is that the subjects in the ABS and Baseline treatments simultaneously view {\itshape both} outcomes and corresponding probabilities of each lottery. As such, on the one hand, the phenomenon at work in the CBS treatment whereby the high-prize attribute receives relatively more attention is less likely to occur, and, on the other hand, the `holistic view' that is enforced in the ABS and Baseline treatments acts as a {\itshape moderator} in the formation of the reference point, in the sense that the other attributes (other than the `high prize') gain relatively more importance. This implies that the range of values that the reference point takes on in the ABS and Baseline treatment is smaller than that of the CBS treatment. In particular, when the lottery-outcome domain comprises gains and losses (experiment 1), the reference point tends to be relatively higher, and when the lottery-outcome domain comprises only gains (experiment 2), the reference point tends to be relatively smaller, compared to that of the CBS treatment. This implies that the subjects assigned to the ABS and Baseline treatments choose in a relatively riskier (resp., safer) way in experiment 1 (resp., experiment 2).

On the basis of the offered interpretation, a key driver of the detected preference reversal is the fact that in the CBS treatment subjects view one outcome at a time. We indeed observe that the preference reversal phenomenon is {\itshape less likely} to occur in the CBS treatment whenever CBS subjects happen to look up {\itshape both} lottery outcomes. In fact, as demonstrated by the combined information-search and choice analysis, while in experiment 1 the probability that the risky lottery is chosen {\itshape increases} when both outcomes are looked up, in experiment 2 the probability that the risky lottery is chosen {\itshape decreases} when both outcomes are looked up. That is, in both experiment 1 and experiment 2, whenever CBS subjects happen to explore both outcomes, their choices tend to be more similar to those of the respective ABS (and Baseline) subjects. We argue that this result provides further support to the proposed interpretation.

We conclude this section by linking our results to the literature. In standard and behavioural models of risky choice, the way the decision-maker allocates attention is a function of the lottery attributes. For example, while - to calculate the expected utility - an expected-utility maximiser looks up all lottery attributes (outcomes and corresponding probabilities) of all lotteries available, a decision-maker using a `maxmin' procedure looks up the lottery outcomes only. On the other hand, while in prospect theory rare events are over-weighted \citep{KahnemanTversky1979}, in salience theory a state of the world $\tilde{s}$ is more salient than some other state $s$ whenever $\tilde{s}$ stands out, i.e., the payoff range of state $s$ is contained in the payoff range of state $\tilde{s}$ \citep{BordaloGennaioliShleifer2012}. In contrast, our findings are in line with the well-documented principle whereby attention is not an output but an {\itshape input of the decision process}, and plays a key role in determining the decision-maker's choice \citep{OrquinMuellerLoose2013}. For example, \citet{PachurSchultemecklenbeckMurphyHertwig2018} identify a systematic relationship between the way in which attention is allocated to the lottery attributes and the prospect theory's preference parameters, such as loss aversion and probability weighting. On the other hand, \citet{HirmasEngelmann2023} demonstrate that imposing a more extensive exposure to a lottery attribute increases its weight on the final decision. We argue that the proposed interpretation of our results is consistent with this strand of literature. 

\section{Conclusion}

In this paper we study the effects of digital nudging by exposing individuals to choice interfaces in a controlled laboratory setting. By suitably modifying the Mouse-Tracing paradigm, we administer three information-search technologies that are based on well-known heuristics, such as alternative-based search (ABS) and characteristic-based search (CBS) \citep{PayneBettmanJohnson1993}, in two pre-registered lottery-choice experiments. Our findings are three-fold. First, we identify an interplay between the heuristic subjects are induced to use and the lottery-outcome domain. In particular, (i) when lottery outcomes comprise gains and losses (experiment 1), inducing subjects to use CBS procedures systematically makes them choose {\itshape safer} lotteries, compared to the subjects that are exposed to the other technologies, and (ii) when lottery outcomes comprise gains only (experiment 2), these results are reversed: inducing subjects to use CBS procedures systematically makes them choose {\itshape riskier} lotteries. Second, we propose an interpretation of our results that is based on prospect theory, whereby, at the core of the matter, is the combined effect on the formation of the reference point of, on the one hand, the heuristic subjects are induced to use and, on the other hand, the lottery-outcome domain. Third, experiment 2 constitutes a replication of our earlier results documenting the effects of exposing individuals to information-search technologies in risky choice.

We conclude by discussing several limitations of this study and outlining potential extensions. First, this paper investigates the effects of exposing individuals to several information-search technologies in two distinct lottery-outcome domains: (i) mixed domain (experiment 1) and (ii) gain domain (experiment 2). A natural follow-up question is to examine the very same research question in a lottery-outcome domain consisting of only losses.

Second, on the basis the information-search and choice analysis, an interpretation of the results has been proposed, according to which there is an interplay between, on the one hand, the heuristic subjects are induced to use, and, on the other hand, the level of attention that - in forming a reference point - subjects pay to the different lottery attributes. It would be interesting to run a replication study by complementing the (instrumental) use of the Mouse-Tracing paradigm with an eye-tracker. Doing so would enable the researcher to refine the information-search analysis, and as a result, conduct a better test of the proposed interpretation.

Third, this paper demonstrates that, in a controlled laboratory setting, administering certain information-search technologies can affect people's risky choices in specific ways. Our results are thus of significant importance for the design and the framing of real-world digital choice platforms. Therefore, a critical follow-up question is to investigate the robustness of the derived results outside the laboratory by testing our hypotheses in the field.


\vspace{1cm}

\addcontentsline{toc}{section}{Bibliography}

\renewcommand{\baselinestretch}{1.0}
\small{

\bibliography{Bibliography}
\bibliographystyle{econometrica}
}



\end{document}